\documentclass[aps,prl,twocolumn,groupedaddress,showpacs]{revtex4}

\usepackage{amsfonts}
\usepackage{amssymb}
\usepackage{amsbsy}
\usepackage{amsmath}
\usepackage{amstext}
\usepackage[mathscr]{eucal}

\usepackage[dvips]{graphicx,color}

\usepackage{verbatim}

\begin{document}

\title{Rotating molecules in optical lattices, alignment and monopole crystals}

\author{J.R. Holt and J.M.F. Gunn}

\affiliation{School of Physics and Astronomy, University of
Birmingham, Edgbaston, Birmingham B15 2TT, U. K.}

\begin{abstract}
The recent progress towards production of near-ground state
quantum-degenerate molecules raises the issue of how such ``small"
molecules behave in an optical lattice. In this Letter we show that
the coupling of the molecular orientation to the local electric
field direction will provide several new phenomena. In the case
where the lasers forming different crystallographic directions of
the lattice are {\em incoherent}, the orientation of the molecules
is conserved (for $L=1$) and a novel form of anisotropic
superfluidity can be expected. When the lasers are {\em coherent},
and the optical lattice is such that the splitting of the rotational
levels is large compared to the centre of mass energies, an
adiabatic description of the molecular orientation is appropriate.
This leads to geometric vector potentials, pseudo-magnetic monopoles
and a frustrated band structure with degenerate minima.
\end{abstract}


\maketitle

The rapid progress in making molecules in ultracold atomic gases
using Feschbach resonances \cite{Regal} and more recently using
photoassociation\cite{Sage,Ye} has provided a new direction for
degenerate gas physics, both in terms of non-equilibrium effects due
to Feschbach sweeps of magnetic fields but also equilibrium many
body phenomena such as the BEC-BCS crossover. The first non-$s$ wave
molecules have been studied in Cs$_2$\cite{GrimmCs} and very
recently in K$_2$\cite{Jin_p}. Novel phenomena such as splitting of
the different multiplet Feshbach resonances has been observed.

It was soon realised that optical lattices could provide a very
sheltered environment for the production of molecules due to control
of occupancy (leading to a lack of harmful three-body collisions).
Initially this led to photoassociative state-selective
production\cite{Rom}. More recently the combination of the
tunability of the optical lattice in terms of tunneling (thorough
laser intensity) and tunability of the interaction through Feshbach
resonances has been used to make molecules with both fermions\cite{
Thal} and bosons\cite{Stof}. The reliability of the occupancy in the
lattice of the molecules is now very good\cite{Volz}.

Very recently Raman down steps (STIRAP)\cite{Wink} towards ground
state molecules have been achieved and with the proposal of
frequency-comb methods\cite{Ye}, it is timely to consider the nature
of near-ground state molecular quantum states in optical lattices.

{\em Large}  molecules (i.e. with an extent large compared to the
optical lattice parameter) have been studied theoretically in
optical lattices by Fedichev et al\cite{Fedi}. The nature of binding
on scales large compared to the optical lattice parameter were
established in some detail. For such a large $s$-wave molecule there
is no coupling to the orientation of the lattice except through the
anisotropy of the center of mass motion.

In this Letter we consider {\it small} molecules where there is a
significant orientational coupling to the optical lattice due to
their inherently anisotropic polarisability. We will see that this
coupling has several pronounced effects on the cenrte of mass
motion, raising issues of the nature of resulting condensed srtates.
(For larger molecules, this presumably corresponds to interference
between the incident optical field and light scattered by one of the
constituent atoms, as experienced by the other atom.)

The most striking effects are in an excited rotational state which
might cause disquiet in terms of lifetimes of such states. However
we take comfort from the case of ortho- and
para-Hydrogen\cite{Silvera,VanK}, where ortho-Hydrogen (which has
$L=1$) has a lifetime at {\em condensed matter densities} of the
order of a day due to the very small matrix elements for relaxation
of the nuclear $I^{\rm tot}=1$ triplet to the singlet.

In fields of the strength\cite{Lee06,RMPmol} occurring in optical
lattices, $L^2$ is a good constant of the motion despite the
manifest lack of spherical symmetry. This is due to the rotational
level spacing being typically six orders of magnitude larger than
the recoil energy. However multiplet {\em splitting} (as against
mixing) already indicates coupling of the orientation to the local
field direction and it is the consequences of such a coupling in an
optical lattice that we will explore in this Letter.

Let the molecule have polarisabilities $\alpha^\parallel$ and
$\alpha^\perp$ parallel and perpendicular to the molecular axis
respectively, and define the average polarisability, ${\overline
\alpha} = \frac{1}{3} (\alpha^\parallel + 2\alpha^\perp)$, and
$\delta\alpha = \alpha^\parallel - \alpha^\perp$. We assume from now
on that the lattice is red-detuned and that $\delta\alpha > 0$. Then
to second order in the applied field, ${\cal E}({\hat {\bf z}})$,
the contribution to the Hamiltonian, ${\cal H}^{(l)}$, for a given
multiplet is:
\begin{equation}{\cal H}^{(l)} = -{\textstyle \frac{1}{4}} {\cal E}^2 \left({\overline \alpha} - {\textstyle
\frac{2}{3}}\delta\alpha\sum_{m=-l}^{m=l}\frac{3m^2-l(l+1)}{(2l+3)(2l-1)}|m\rangle\langle
m|\right)\end{equation} (see, for example ref (\cite{Mizu}) where
the angular momentum quantum numbers are $l$ and $m$ with the axis
of quantisation parallel to the field.

For a one-dimensional position-dependent electric field, ${\cal
E}(X){\hat {\bf z}}$, where $X$ is the centre of mass position of
the molecule, we define  (now considering $L=1$):
\begin{equation}{\cal H}_z(X)=-{\textstyle \frac{1}{4}} {\cal E}^2(X)
\left(\alpha_1 +\alpha_2(m)|m_z=0\rangle\langle m_z=0|\right)
\label{eq:incoh}\end{equation} where $z$ denotes the direction of
the field, $\alpha_1 = {\overline \alpha}- \frac{2}{15}
\delta\alpha$ and $\alpha_2 =\frac{2}{5} \delta\alpha$. The physical
interpretation of this is semiclassical. For the states $m=\pm 1$
the molecule presents its less polarisable directions to the field.
This means the barriers to tunneling are reduced compared to random
orientation but also the minima are raised compared to random
orientation. $m=0$ presents (semi-classically) the most polarisable
axis half the time and hence has higher barriers, but lower minima.

We see that in one dimension the orientation of the molecule is a
constant of the motion. It is perhaps more surprising that we can
make statements of a similar nature in higher dimensions. However
the mutual coherence or incoherence of the lasers in the different
axes of the optical lattice affects the results substantially.

In the {\em incoherent} case, let the polarisation of the electric
field of the laser associated with optical lattice axes $i=x$, $y$
or $z$ be denoted by the unit vectors ${\hat{\bf e}}_i$. Then the
perturbed energy is the sum of three terms of the form
(\ref{eq:incoh}):
\begin{eqnarray}{\cal H}({\bf R})\!\!&=&\!\!\! {\cal H}_{{\hat{\bf
e}}_x}(X) + {\cal H}_{{\hat{\bf e}}_y}(Y)+{\cal
H}_{{\hat{\bf e}}_z}(Z) \nonumber\\
&=&\!\!\! -{\textstyle \frac{1}{4}} \sum_{i=x,y,z}\!\!\! {\cal
E}^2(R_i) \left( \alpha_1 +\alpha_2|m_i =0\rangle\langle
m_i=0|\right) \label{eq:threed}\end{eqnarray} where $m_i$ is an
eigenvalue of ${\hat{\bf e}}_i\cdot{\bf L} $.

The result is particularly straightforward if the set $\{{\hat{\bf
e}}_i \}$ is orthogonal. Then the set $\{ |m_i =0\rangle\}$ (the
``polar basis") is also orthogonal. In that case the three
contributions to the Hamiltonian from the projection operators $|m_i
=0\rangle\langle m_i=0|$ commute and hence the Hamiltonian is
diagonal. A molecule in one of the eigenstates, $|m_i =0\rangle$,
will experience two directions with small periodic potentials and in
the direction parallel to ${\hat{\bf e}}_i$, a deeper one. For
$L>1$, the situation is more complicated: for example it is not
guaranteed that the $m=0$ states along three axes are orthogonal
(e.g. $L=2$), so the projection operators do not commute.

The basis set allows some translation of the $S=1$ spinor results
from references\cite{Spinor}. The molecule-molecule scattering will
have singlet and quintuplet scattering lengths which determine
whether the ground state is a polar state (as in the $S=1$ case)
which is compatible with the diagonalised Hamiltonian, albeit with
anisotropic superfluid properties. It is less clear what happens for
the case which would be ferromagnetic without the lattice.

Let us now turn to the {\em coherent} case. Here the polarisation of
the optical lattice is position-dependent. For definiteness, pick
\begin{equation}{\bf E}({\bf R}) = {\cal E}_0\left({\hat{\bf x}}\sin
kZ+{\hat{\bf y}}\sin kX+{\hat{\bf z}}\sin kY\right)\cos \Omega
t\label{eq:efield}\end{equation} and for future reference define
$|{\bf E}({\bf R})| = {\cal E} ({\bf R})$. Then the antinodes, i.e.
the minima of the optical lattice potential, are at ${\bf R} =
(\lambda/4)(2\ell+1,2m+1,2n+1)$. There are four distinct minima (as
the Hamiltonian does not distinguish the sign of the electric
field), due to the relative signs of the components of ${\bf E}$.

The case where there is an adiabatic separation of the energy scales
of splitting the angular momentum multiplet and the centre of mass
motion in one of the optical lattice wells (and hence in an even
more pronounced form in terms of tunneling) will be the focus for
the remaining part of this Letter. In that limit the molecule
preserves its orientational state relative to the local electric
field direction, determined by the centre of mass position. Unlike
the incoherent case the results are qualitatively similar for all
$L>0$. Berry and Robbins\cite{Berry1} studied a related problem
where there was a {\em linear} coupling between angular momentum and
spatial variables in an adiabatic limit which we will comment on
presently.

We form the adiabatic basis, $|m, {\bf R}\rangle$, by rotating the
reference basis set, $|m\rangle$, so that the $z$-axis is
transformed to the local electric field direction appropriate to
centre of mass position $\bf R$: \begin{eqnarray}|m, {\bf
R}\rangle &=& D(-\eta({\bf R}),-\zeta({\bf R}),0)|m\rangle\nonumber\\
&=& \exp ({\rm i}\eta({\bf R})\sigma_z) \exp ({\rm i}\zeta({\bf R})
\sigma_y)|m\rangle\nonumber\end{eqnarray} where $\{\sigma_i\}$, with
$i=x$, $y$ or $z$, are the spin-1 Pauli matrices. Thus
$${\hat  {\bf
E}}({\bf R}) \cdot {\bf L}|m, {\bf R}\rangle = m| m, {\bf R}\rangle
$$ with ${\hat {\bf E}}({\bf R})$ is the unit vector corresponding
to the local electric field.

We then look for an approximation to the solution, $\Psi ({\bf R},
\zeta, \eta)$, of the Schrodinger equation for the molecule
$$\left[-\frac{\hbar^2}{2M} \nabla^2_{\bf R} -{\textstyle
\frac{1}{4}} {\cal E}^2({\bf R}) \left( \alpha_1 +\alpha_2|0, {\bf
R}\rangle\langle 0, {\bf R}|\right)\right] \Psi = \epsilon \Psi$$

The adiabatic approximation is to assume $$\Psi({\bf R}, \zeta,
\eta) \simeq \sum_{m=-1}^1 \psi_m({\bf R})|m, {\bf R}\rangle$$ which
leads (upon substitution and performing an integration over the
angular variables) to the effective Schrodinger equation in terms of
a vector potential\cite{Berry} ${\bf A}^{nl}=-{\rm i} \langle n,
{\bf R}|\nabla_{\bf R}|l, {\bf R}\rangle$ :
\begin{equation}\left( -\frac{\hbar^2}{2m}{\bf
\Delta}^{nl}\cdot {\bf \Delta}^{lm}  +V^{nm}({\bf
R})\right)\psi^m({\bf R})=E\psi^n({\bf
R})\label{eq:Sch}\end{equation} where we define the covariant
derivative, ${\bf \Delta}^{nl}= \delta^{nl}\nabla_{\bf R} -{\rm
i}{\bf A}^{nl}$ and $$V^{nm}({\bf R})=-{\textstyle\frac{1}{4}}{\cal
E}^2({\bf R})\left(\alpha_1 \delta^{nm}\!\!+ \alpha_2
P^{nm}_0\right) + \sum_{l\ne n}\!\!{\bf A}^{nl}\cdot{\bf
A}^{lm}\delta^{nm}$$ (with no summation convention), where $P^{nm}_0
= \delta^{m0}\delta^{n0}$.
 Non-Abelian vector potentials lead to effects experienced by
multi-level atoms in light fields: the case of spin-orbit
effects\cite{Car}, laser-assisted tunneling\cite{Oster} and
tripod-field induced degenerate dark states\cite{Ruse}. This Letter
contains the first example the new phenomena experienced in a {\em
three}-dimensional extended setting with such non-Abelian fields.

We may evaluate the vector potential:
$${\bf A}^{nl} = \left[(\sigma^{nl}_z\cos \zeta
 + \sigma^{nl}_x\sin \zeta )\nabla_{\bf R} \eta + \sigma^{nl}_y\nabla_{\bf R}
\zeta \right]$$ and hence determine the associated field:
\begin{eqnarray}F^{mn}_{ij}&=&\partial_iA^{mn}_j- \partial_j A^{mn}_i -
{\rm i}[A_i,A_j]^{mn}\nonumber\\
&=& 2(\partial_i\eta \partial_j\zeta - \partial_j\eta
\partial_i\zeta)( \sigma^{mn}_x\cos \zeta - \sigma^{mn}_z\sin \zeta
)\nonumber\end{eqnarray} implying that the pseudo -field, $B_i^{mn}=
\frac{1}{2} \epsilon_{ijk} F_{jk}^{mn}$ is
\begin{equation}{\bf B}^{mn} = 2 \nabla \zeta \times \nabla \eta( \sigma^{mn}_x\cos \zeta - \sigma^{mn}_z\sin
\zeta)\label{eq:eul}\end{equation} with the polar angles, $\zeta$
and $\eta$, playing the role of ``Euler potentials" for the field
$\bf B$.

An immediate consequence of equation (\ref{eq:eul}) is that $\nabla
\cdot {\bf B} =0$ except possibly at singularities of $\nabla \eta$
and/or $\nabla \zeta$, which occur at the nodes of the electric
field of the optical lattice (\ref{eq:efield}). Firstly consider the
form of the electric field in the vicinity of the node at the origin
$${\bf E}({\bf R}) \simeq k{\cal E}\left( {\bf {\hat x}}Z +{\bf {\hat y}}X
+{\bf {\hat z}}Y \right)$$ To examine the nature of the
pseudomagnetic field and its flux, it is easiest to calculate
(\ref{eq:eul}) in polar coordinates around the node. The relation
between the polar coordinates of the Electric field vector,
$(\zeta,\eta)$, and those of the centre of mass $(\theta,\phi)$ are
determined implicitly from:
\begin{eqnarray}
{\hat E}_x &=& \sin \zeta \cos \eta = Z/R = \cos\theta\nonumber \\
{\hat E}_y &=& \sin \zeta \sin \eta = X/R = \sin\theta \cos\phi\nonumber\\
{\hat E}_z &=& \cos \zeta \phantom{\sin \eta}= Y/R = \sin\theta
\sin\phi\nonumber
\end{eqnarray}
Using expressions for $\cos \zeta$ and $\tan\eta$ we find after some
algebra that:
$$\nabla \zeta \times \nabla \eta = - \frac{{\hat {\bf R}}}{R^2}
\frac{1}{\sin\zeta}$$ and hence \begin{equation}{\bf B}^{mn}({\bf
R}) = 2\frac{{\hat {\bf R}}}{R^2} (\sigma_z^{mn} -
\sigma_x^{mn}\thinspace\cot \zeta)\label{eq:bfield}\end{equation} in
the vicinity of the node.

We will now calculate the flux through a spherical surface of ${\bf
B}^{mn}$. We will neglect the Dirac-like strings which emerge along
the positive and negative $y$-axes (reminiscent of the two strings
in Schwinger's treatment\cite{Sch} of monopoles), as the strings are
irrelevant to our final lattice treatment.

The second term in Eqn. (\ref{eq:bfield}) vanishes as there is a
cancelation from the contributions from $0\le \phi < \pi$ and $\pi
\le \phi < 2\pi$. The first term provides $8\pi \sigma_z$ flux
through the surface, corresponding to monopoles of charge $\pm8\pi$.
These are the analogues of the monopole in the the work of Berry and
Robbins\cite{Berry1}.

The field in the vicinity of the nodes at ${\bf R}=(m,n,\ell)$ has
the form:
$${\bf E}({\bf R}) \propto (-1)^\ell(Z-\ell){\hat{\bf x}} +
(-1)^m(X-m){\hat{\bf y}}+(-1)^n(Y-n){\hat{\bf z}}$$ In the case of
one or three of $\ell$, $m$ and $n$ being odd, then either the sign
of $\nabla \eta$ or $\nabla \zeta$ reversed. Thus the ($m=1$)
monopole has negative charge. Conversely, if there are zero or two
odd integers, the ($m=1$) monopole has a positive charge as neither
or both gradients are reversed.

Thus the monopoles reside in a Na-Cl lattice, coincident with the
nodal points of the electric field, of alternating positive and
negative magnetic charges. The signs of all charges are reversed if
the angular momentum of the molecule is reversed.

Since the adiabatic limit will require a strong optical lattice it
is natural to construct the tight binding model corresponding to the
continuum Schrodinger equation, Eqn. (\ref{eq:Sch}). The lattice is
cubic and in the adiabatic limit the distinction between the field
directions on the different sites in the unit cell is irrelevant as
they are all perfectly adiabatically connected and degenerate in
energy. The monopoles reside on the dual lattice, with the opposite
signs adopting a Na-Cl structure. We will focus on the $m=1$ case.

\begin{figure}[h]
\includegraphics[width=0.3\textwidth,clip=true,angle=-90]{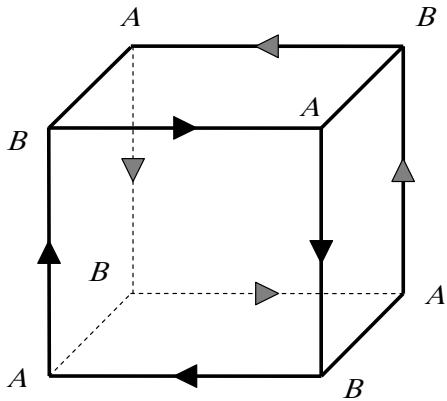}
\caption{\label{fig:cuboid2} The choice of Peierls factors in Eqn.
(\ref{eq:Pei}). An arrow from $B$ to $A$, for example, indicates a
factor of ${\rm e}^{{\rm i}2\pi/3}$ in the term hopping from $B$ to
$A$.}
\end{figure}

We must assign Peierls factors, ${\rm e}^{{\rm i}A_{{\bf n},{\bf
n'}}}$, to the hopping term on each link between sites ${\bf n}$ and
${\bf n'}$. The flux, $\Phi_{{\bf n}',{\bf n}'',{\bf n}''',{\bf
n}''''}$, through a plaquette is counted modulo $2\pi$ as ${\rm
e}^{{\rm i}\Phi} = {\rm e}^{{\rm i} \sum_\Box A}$. Then  we must
pick $A_{{\bf n},{\bf n'}}$ so that we get a flux of $\pm 2\pi/3$
through each plaquette, so we obtain a total flux out of the six
plaquettes surrounding a monopole of $\mp 8\pi$. The Peierls factors
are chosen as indicated in Fig. (\ref{fig:cuboid2}), where we see
there are two sites per unit cell denoted $A$ and $B$. This leads to
the secular determinant (here $t>0$ and $\omega={\rm e}^{{\rm
i}2\pi/3}$):
\begin{equation} \left| \begin{array}{cc}
-\epsilon & -2t(\omega X + Y + \omega^{-1} Z) \\
-2t(\omega^{-1} X +  Y + \omega Z) & -\epsilon
\end{array} \right| =0\label{eq:Pei}\end{equation} where: $X=\cos k_x$ and
similarly for $Y$ and $Z$. This implies
\begin{equation}\epsilon_\pm({\bf k}) = \pm \sqrt{2}\thinspace t
\sqrt{(X-Y)^2+(Y-Z)^2+(Z-X)^2}\end{equation} The two bands touch
along the $(\pm 1,\pm 1,\pm 1)$ directions, with the dispersion
relation in that vicinity (defining $k^\parallel$ parallel to the
$(1,1,1)$ direction and $k^\perp\thinspace (\ge 0)$ radially
perpendicular to the direction) being
$\epsilon_\pm(k^\parallel,k^\perp)\simeq \pm\sqrt{3} |\sin
(k^\parallel/\sqrt{3})| k^\perp$ for $k^\perp/k^\parallel \ll 1$.

The most important aspect of the dispersion relations relates to the
frustrated nature of the Hamiltonian. The minima of $\epsilon_-({\bf
k})$ are at the three points in the Brillouin Zone: ${\bf k}=
(0,0,\pi)$ and cubic equivalents. Because the (direct) lattice is
FCC, the other candidates ${\bf k}= (0,\pi,\pi)$ are outside the
first Brillouin Zone and equivalent to the first set by adding a
reciprocal lattice vector, $\pi(1,1,1)$, of the BCC reciprocal
lattice. Thus the minima in energy are $-4t$ which is higher than
the $-6t$ expected of a simple cubic system. The wavefunctions on
the $A$ and $B$ sublattices, $(\psi_A,\psi_B)$, for the three minima
are: for ${\bf k}^{(Z)}= (0,0,\pi)$,
$(\psi_A^{(Z)},\psi_B^{(Z)})=(1,-\omega)$; for ${\bf k}^{(Y)}=
(0,\pi,0)$, $(\psi_A^{(Y)},\psi_B^{(Y)})=(1,-1)$; for ${\bf
k}^{(X)}= (\pi,0,0)$, $(\psi_A^{(X)},\psi_B^{(X)}
)=(1,-\omega^{-1})$.

The three minima, in conjunction with the possible competition of
singlet and quintuplet scattering lengths, make the nature of the
mean field ground state in general unlcear.

In conclusion we have shown that in both incoherent and coherent
optical lattices the orientation of a molecule with $L\ne 0$ couples
to the direction of the optical field. In the incoherent case the
orientation is conserved in a highly symmetric manner for $L=1$. For
the coherent case the strong optical fields lead to the angular
momentum around the local electric field being an adiabatic
invariant. The consequences of the adiabatic approximation failing
at the nodal points of the lattice, in the classically disallowed
region, affects the tunneling of the molecule. This leads to
pseudomagnetic monopoles with their flux influencing the band
structure of the molecule in a striking manner. The consequences for
condensation remain to be explored, as does the configuration of
monopoles in cases of general (eg non-bipartite) dual latttices.

JRH would like to thank EPSRC for a studentship and the University
of Birmingham for support. JMFG would like to thank Nigel Cooper,
Martin Long and Nicola Wilkin for several helpful discussions and
KITP, University of California at Santa Barbara,  for hospitality
while some of the work was performed. This work was also supported
by EPSRC grant GR/R00920.

\end{document}